\documentclass[twocolumn,prb,aps,superbib,tightenlines,floatfix,superscriptaddress,showpacs,letter]{revtex4}
\usepackage{amsmath}
\usepackage{graphicx}
\usepackage{amssymb}

\begin{document}

\title{Kinetics of Spin Relaxation in Wires and Channels: Boundary Spin Echo and Tachyons}

\author{Valeriy A. Slipko}
\affiliation{Department of Physics and Astronomy and USC
Nanocenter, University of South Carolina, Columbia, SC 29208, USA}
\affiliation{ Department of Physics and Technology, V. N. Karazin
Kharkov National University, Kharkov 61077, Ukraine }

\author{Yuriy V. Pershin}
\email{pershin@physics.sc.edu}

\affiliation{Department of Physics and Astronomy and USC
Nanocenter, University of South Carolina, Columbia, SC 29208, USA}

\begin{abstract}
In this paper we use a spin kinetic equation to study spin polarization dynamics in 1D wires and 2D channels. This approach is valid in both diffusive and ballistic spin transport regimes and, therefore, more general than the usual spin drift-diffusion equations. In particular, we demonstrate that in infinite 1D wires with Rashba spin-orbit interaction the exponential spin relaxation decay can be modulated by an oscillating function. In the case of spin relaxation in finite length 1D wires, it is shown that an initially homogeneous spin polarization spontaneously transforms into a persistent spin helix. An interesting sound waves echo-like behavior of initially localized spin polarization packet is found in finite length wires. We show that a propagating spin polarization profile reflects from a system boundary and returns back to its initial position similarly to the reflectance of sound waves from an obstacle. Green's function of spin kinetic equation is found for both finite and infinite 1D systems. Moreover, we demonstrate explicitly that the spin relaxation in 2D channels with Rashba and Dresselhaus spin-orbit interactions of equal strength  occurs similarly to that in 1D wires of finite length. Finally, a simple transformation mapping 1D spin kinetic equation into the Klein-Gordon equation with an imaginary mass is found thus establishing an interesting connection between semiconductor spintronics and relativistic quantum mechanics.
\end{abstract}

\pacs{72.15.Lh, 72.25.Dc, 85.75.2d, 14.80.-j} \maketitle

\section{Introduction}

Dynamics of electron spin polarization in semiconductor structures~\cite{Wu10a} has attracted a lot of attention recently in the context of spintronics~\cite{Zutic04a}.  Typically, the most important mechanism of spin relaxation in semiconductors lacking inversion symmetry is the D'yakonov-Perel' spin relaxation mechanism~\cite{Dyakonov72a,Dyakonov86a}, intimately related to the spin splitting of the electronic states. Numerous studies of D'yakonov-Perel' spin relaxation in the past have dominantly concentrated on spin relaxation in infinite two-dimensional (2D) systems~\cite{Dyakonov72a,Dyakonov86a,Sherman03a,Burkov04a,Saikin04a,Pershin04b,Pershin05a,Lyubinskiy06b,Pershin07a,Weng08a,Kleinert09a,pershin10a,Tokatly10a,Tokatly10b}, while
the electron spin relaxation in confined geometries is not yet well understood.
This problem, however, is of crucial importance since actual spin-based electronic devices \cite{Bandy09a} are normally of a finite size.

There are only several examples in the literature where the effects of boundary conditions on D'yakonov-Perel' spin relaxation have been explored theoretically and/or experimentally.
In particular, investigations of spin relaxation in systems with boundaries include spin relaxation in 2D channels \cite{Kiselev00a,Schwab06a,Holleitner07a,Chang09a,Frolov10a,Liu10a}, 2D half-space \cite{Pershin05c}, 2D systems with antidots \cite{Pershin04a}, large quantum dots \cite{Lyubinskiy06a,Koop08a}, and one-dimensional (1D) finite-length wires \cite{Slipko11a} and rings \cite{Slipko11b}. The common conclusion of these studies is that in the diffusive regime of spin transport the introduction of boundary conditions results in an increased spin lifetime. In particular, in Ref. \onlinecite{Slipko11a} the present authors studied spin relaxation in finite length quantum wires using drift-diffusion equations approach. It was demonstrated that a persistent spin helix \cite{Schliemann03a,Bernevig06a} spontaneously emerges in course of relaxation of homogeneous spin polarization.

In this paper, we extend the results of our previous study~\cite{Slipko11a} applying spin kinetic equation approach to the problem of spin relaxation in 1D wires and 2D channels with spin-orbit interaction. The formalism of spin kinetic equation describes both diffusive and ballistic regimes of spin transport at any value of spin rotation angle per mean free path. Therefore, the spin kinetic equation is more general than traditional spin drift-diffusion equations \cite{Yu02a,Burkov04a,Saikin04a,Pershin04b,arXiv_1007_0853v1} and can be strictly justified by using quasiclassical Green's functions \cite{Raimondi06a,Schwab06a}. Basically, the spin kinetic equation describes the spin polarization dynamics on shorter time and space scales than those of spin drift-diffusion equations. Consequently, the application of spin kinetic equation to the problem of spin relaxation provides new insights into the spin polarization dynamics. In particular, in this paper we report {\it the boundary spin echo effect} in which a localized spin polarization profile reflects from a sample boundary and returns with a decreased amplitude to its initial position similarly to the usual spin echo of sounds waves. This type of spin echo is essentially different from the spin echo in nuclear magnetic resonance. The later is related to nuclear spins dephasing in an applied magnetic field \cite{Abragam83}.  Moreover, we find a simple transformation that maps the spin kinetic equation into the Klein-Gordon equation with an imaginary mass, which is a relativistic analog of the Schr\"odinger's equation. It is worth mentioning that the Klein-Gordon equation with an imaginary mass describes {\it tachyons}~\cite{Feinberg67a,Murphy72a,Recami86a,Dawe92a} - hypothesized particles that travel faster than light. Consequently, we suggest that semiconductor spintronics structures have a potential to be used as a laboratory test bed for relativistic quantum mechanics.

Our paper is organized as follows. In Sec. \ref{sec2} we introduce a spin kinetic equation and derive a set of equations for spin polarization components and boundary conditions. We also demonstrate that the spin kinetic equation can be transformed into the Klein-Gordon equation. Next, in Sec. \ref{sec3}, we employ the spin kinetic equation to study spin relaxation in infinite and finite length wires with Rashba spin-orbit interaction. In particular, in Sec. \ref{sec34}, we present the boundary spin echo effect. The spin relaxation in two-dimensional channels with Rashba and Dresselhaus~\cite{Dresselhaus55a} spin-orbit interactions of equal strength is investigated in Sec. \ref{sec4}. We demonstrate that such a problem, for channels in a specific direction, can be mapped into the problem of spin relaxation in finite length wires with Rashba interaction only. The results of our investigations are summarized in Sec. \ref{sec5}.

\section{Theoretical Framework} \label{sec2}
\subsection{Spin kinetic equation}

The main goal of this paper is to study the kinetics of spin relaxation in
wires made of 2D quantum well or heterostructure with Rashba spin-orbit interaction~\cite{Bychkov84a}.
Therefore, we first introduce the Hamiltonian for an electron in
2D space in the presence of spin-orbit interaction and all important
parameters that will be used later in the spin kinetic equation.
The standard Hamiltonian with the Rashba~\cite{Bychkov84a} term
is given by
\begin{equation}
 H= H_0+ H_R=\frac{\mathbf{ p}^2}{2m}+\alpha\left(\mathbf{\sigma}
\times\mathbf{ p}\right)\cdot\mathbf{z},
\label{ham}
\end{equation}
where $\mathbf{{p}}=( p_x, p_y)$ is the 2D electron
momentum operator, $m$ is the effective electron's mass,
$\mathbf{\sigma}$ is the Pauli-matrix vector, $\alpha$ is the
spin-orbit coupling constant and $\mathbf{z}$ is a unit vector
perpendicular to the confinement plane.

It is not difficult to demonstrate~\cite{arXiv_1007_0853v1} that in the case of Hamiltonian
(\ref{ham}) the quantum mechanical evolution of a spin of an
electron with a momentum $\mathbf{p}$ can be reduced to a spin
rotation with the angular velocity $\Omega=2\alpha p/\hbar$ about
the axis determined by the unit vector
$\mathbf{n}=\mathbf{p}\times \mathbf{z}/p$. In this way, the spin-orbit coupling constant
$\alpha$ enters into equations through the parameter $\eta=2\alpha
m \hbar^{-1}$, which gives the spin precession angle per unit
length.

Besides this evolution, 2D electrons experience different bulk
scattering events such as, for example, due to phonons or
impurities. These scatterings randomize the electron trajectories.
Correspondingly, the direction of spin rotation becomes
fluctuating what causes average spin relaxation (dephasing). This
is the famous D'yakonov-Perel' spin relaxation
mechanism.~\cite{Dyakonov72a,Dyakonov86a} The time scale of the
bulk scattering events can then be characterized by a single rate
parameter, the momentum relaxation time $\tau$. It is connected to
the mean free path by $\ell=v \tau$, where $v=p/m$ is the mean
electron velocity. To take into account  these scatterings we use
a kinetic model of spin transport presented below. When a characteristic system size $L\gg \ell$ and the time scale of interest is much longer than $\tau$ ,
 the spin kinetic model yields the spin
drift-diffusion equations (such as, e.g., reported in Ref. \onlinecite{arXiv_1007_0853v1}).

In the semi-classic approximation the kinetic equation for electron spin polarization  can be written as (see, e.g., Refs. \onlinecite{Lyubinskiy06a}, \onlinecite{Lyubinskiy06b}, \onlinecite{Schwab06a}, \onlinecite{Raimondi06a})
\begin{eqnarray}
\left(\frac{\partial }{\partial t}+\frac{\mathbf{p}}{m}\cdot\nabla\right)
\mathbf{S_p}=\mathbf{\Omega_p}\times\mathbf{S_p}+St\{\mathbf{S_p}\},
\label{KinEq}
\end{eqnarray}
where $\mathbf{S_p}(\mathbf{r},t)$ is the vector of spin polarization of electrons moving with momentum $\mathbf{p}$, and $St\{\mathbf{S_p}\}$ is the collision integral describing electron scattering processes. In the $\tau$-approximation  the collision integral is given by
\begin{eqnarray}
St\{\mathbf{S_p}\}=-\frac{1}{\tau}(\mathbf{S_p}-\langle\mathbf{S_p}\rangle),
\label{CollisionIntegral}
\end{eqnarray}
where the angle brackets denote averaging over direction of electron momentum.
The conditions of applicability of Eq. (\ref{KinEq}) can be found in Ref. \onlinecite{Schwab06a}.
The collision integral (\ref{CollisionIntegral}) corresponds to the elastic scattering of electrons by strong scatterers with a characteristic time $\tau$ between the collisions.
Note that the collision integral (\ref{CollisionIntegral}) conserves the total spin polarization redistributing spin polarization between electrons moving in different directions.
For 1D case the average spin polarization simplifies to the following expression $\langle\mathbf{S_p}\rangle=(\mathbf{S}^+ +\mathbf{S}^-)/2 $, where $\mathbf{S}^+$ and $\mathbf{S}^-$ are the spin polarizations of electrons  moving along the wire in the positive (with momentum $\mathbf{p}=mv\mathbf{e}_x$), and negative ($\mathbf{p}=-mv\mathbf{e}_x$) directions with the average velocity $v$. Thus, the kinetic equation (\ref{KinEq}) for 1D wire takes the form of the system of two vector equations
\begin{eqnarray}
\left(\frac{\partial }{\partial t}+v\frac{\partial }{\partial x}\right)
\mathbf{S}^+=-\Omega\mathbf{e}_y\times\mathbf{S}^+ -\frac{1}{2\tau}(\mathbf{S}^+-\mathbf{S}^-),
\label{KinEqPlus}  \\
\left(\frac{\partial }{\partial t}-v\frac{\partial }{\partial x}\right)
\mathbf{S}^-=\Omega\mathbf{e}_y\times\mathbf{S}^- -\frac{1}{2\tau}(\mathbf{S}^--\mathbf{S}^+).
\label{KinEqMinus}
\end{eqnarray}
This system of equations should be complimented by initial conditions for spin densities
$\mathbf{S}^{+}$ and $\mathbf{S}^{-}$
\begin{eqnarray}
\mathbf{S}^+(x,t=0)=\mathbf{S}^+_0(x),
\label{ICSp} \\
\mathbf{S}^-(x,t=0)=\mathbf{S}^-_0(x),
\label{ICSm}
\end{eqnarray}
and by boundary conditions for finite length wires. The boundary conditions conserving the spin polarization
in elastic scatterings at the boundary $\Gamma=[x=0,x=L]$ have the form
\begin{eqnarray}
(\mathbf{S}^+-\mathbf{S}^-)|_{\Gamma}=0.
\label{BCpm}
\end{eqnarray}

Taking the sum and difference of Eqs. (\ref{KinEqPlus}, \ref{KinEqMinus}) we easily obtain
\begin{eqnarray}
\frac{\partial \mathbf{S}}{\partial t}
+v\frac{\partial \mathbf{\Delta}}{\partial x}
+
\Omega\mathbf{e}_y\times\mathbf{\Delta}=0,
\label{KinEqS}  \\
\frac{\partial \mathbf{\Delta}}{\partial t}
+v\frac{\partial \mathbf{S}}{\partial x}
+
\Omega\mathbf{e}_y\times\mathbf{S}+\frac{\mathbf{\Delta}}{\tau}=0
\label{KinEqDelta},
\end{eqnarray}
where the following notations are used: $\mathbf{S}=\mathbf{S}^++\mathbf{S}^-$ and  $\mathbf{\Delta}=\mathbf{S}^+-\mathbf{S}^-$.
As we are mainly interested  in finding the total spin polarization $\mathbf{S}$, $\mathbf{\Delta}$ can be eliminated from Eqs. (\ref{KinEqS}),  (\ref{KinEqDelta}) via the following transformation. First of all, we multiply Eq. (\ref{KinEqDelta}) by $e^{t/\tau}$ and rewrite it as
\begin{equation}
\frac{\partial \left( e^{t/ \tau} \mathbf{\Delta}\right) }{\partial t}
+ve^{t/ \tau}\frac{\partial \mathbf{S}}{\partial x}
+ \Omega e^{t/ \tau}\mathbf{e}_y\times\mathbf{S}=0 \label{eq1}.
\end{equation}
Then, Eq. (\ref{KinEqS}) is multiplied by $e^{t/\tau}$ and is differentiated with respect to time
\begin{equation}
\frac{\partial}{\partial t}\left( e^{t/ \tau}  \frac{\partial \mathbf{S}}{\partial t} \right)
+v\frac{\partial }{\partial x} \frac{\partial \left( e^{t/ \tau} \mathbf{\Delta}\right) }{\partial t}
+ \Omega\mathbf{e}_y\times \frac{\partial \left( e^{t/ \tau} \mathbf{\Delta}\right) }{\partial t}=0 \label{eq2}.
\end{equation}
Finally, we substitute $\partial \left( e^{t/ \tau} \mathbf{\Delta}\right) / \partial t$ from Eq. (\ref{eq1}) into Eq. (\ref{eq2}). The resulting
equations for spin polarization  can be presented as
\begin{eqnarray}
\frac{\partial^2 \mathbf{S}}{\partial t^2}+\frac{1}{\tau}\frac{\partial \mathbf{S}}{\partial t}
-v^2\frac{\partial^2 \mathbf{S}}{\partial x^2}&-&2\Omega v \mathbf{e}_y\times\frac{\partial\mathbf{S}}{\partial x} \nonumber \\
&+&\Omega^2\left(\mathbf{S}-S_y\mathbf{e}_y \right)=0,
\label{KinEqSFin}
\end{eqnarray}
where $S_y$ is the $y$-component of $\mathbf{S}$.

Next, we would like to reformulate the boundary condition given by Eq. (\ref{BCpm}) in terms of the function $\mathbf{S}$ only. It is easy to notice that Eq. (\ref{BCpm})
corresponds to $\left.\mathbf{\Delta}\right|_\Gamma=0$. Substituting this boundary value of $\mathbf{\Delta}$ into Eq. (\ref{KinEqDelta}) we obtain
\begin{eqnarray}
\left.\left(v\frac{\partial \mathbf{S}}{\partial x}
+\Omega\mathbf{e}_y\times\mathbf{S}\right)\right|_\Gamma=0.
\label{BCS}
\end{eqnarray}
Previously, the same form of boundary condition was derived using the Green's function method \cite{Galitski06a}.

Moreover, since Eq. (\ref{KinEqSFin}) is the second order differential equation with respect to time, one must specify both the spin polarization and its time derivative at the initial moment of time $t=0$
\begin{eqnarray}
\mathbf{S}(x,t=0)=\mathbf{S}_0(x),~~\left(\frac{\partial\mathbf{S}}{\partial t}\right)_{t=0}=\dot{\mathbf{S}}_0(x).
\label{ICS}
\end{eqnarray}
Note that if we know $\mathbf{\Delta}$ at the initial moment of time $t=0$ then we can find the first time derivative of $\mathbf{S}$ at $t=0$ using Eq. (\ref{KinEqS}). In particular, if $\Delta(x,t=0)=0$ then it follows from Eq. (\ref{KinEqS}) that $\dot{\mathbf{S}}_0(x)=0$. 

Considering $y$-components of Eqs. (\ref{KinEqSFin})-(\ref{BCS}) we find that $S_y$ is not coupled to any other component of spin polarization. Specifically, Eqs. (\ref{KinEqSFin})-(\ref{BCS}) for
$S_y$ can be rewritten as
\begin{eqnarray}
\frac{\partial^2 S_y}{\partial t^2}+\frac{1}{\tau}\frac{\partial S_y}{\partial t}
-v^2\frac{\partial^2 S_y}{\partial x^2}=0,
\label{KinEqSy}\\
\left.\frac{\partial S_y}{\partial x}\right|_\Gamma=0.
\label{BCSy}
\end{eqnarray}
Consequently, selecting $S_y(x,t=0)=0$ we can safely take out $S_y$ from our consideration.

Let us introduce a complex polarization
\begin{eqnarray}
S=S_x+iS_z.
\label{ComplexS}
\end{eqnarray}
It is straightforward to show
that Eq. (\ref{KinEqSFin}), and boundary conditions (\ref{BCS})
can be rewritten in a more compact form using $S$:
\begin{eqnarray}
\frac{\partial^2 S}{\partial t^2}+\frac{1}{\tau}\frac{\partial S}{\partial t}
-\left(v\frac{\partial}{\partial x}-i\Omega\right)^2 S=0,
\label{ComplexSEq} \\
\left. \left(v\frac{\partial S}{\partial x}-i\Omega S\right)\right|_\Gamma=0.
 \label{ComplexSBC}
\end{eqnarray}
Moreover, it follows from Eqs. (\ref{ComplexSEq}) and (\ref{ComplexSBC}) that they have only one stationary solution
\begin{eqnarray}
S=S_ 0e^{i\eta x}
\label{ComplexSHW},
\end{eqnarray}
where $S_0$ is an arbitrary complex constant.
This solution is so-called  spin helix \cite{Pershin05a,Bernevig06a}, which is persistent in 1D Rashba wires despite electron collisions.
Taking the real and imaginary parts of Eq. (\ref{ComplexSHW}) with $S_0=Ae^{i\delta}$, we find the usual representation of spin helix
\begin{eqnarray}
S_x=A\cos(\eta x+\delta), ~S_z=A\sin(\eta x+\delta) \label{SHW},
\end{eqnarray}
where $A$ is the spin helix amplitude, and $\delta$ is its phase. Since the solution (\ref{SHW}) of Eqs. (\ref{ComplexSEq}) and (\ref{ComplexSBC}) is the only stationary one, any initial spin
polarization distribution eventually transforms into the spin helix~\cite{Slipko11a}, in some cases, however,
having zero amplitude $A=0$.

\subsection{Klein-Gordon equation}
We can exclude rotations of spin polarization vector that are still present in Eq. (\ref{ComplexSEq}) and Eq. (\ref{ComplexSBC}) by introducing a complex field $u$ via
\begin{eqnarray}
u(x,t)=e^{-i\eta x}e^{\frac{t}{2\tau}}S(x,t) \label{u}.
\end{eqnarray}
 It can be shown that Eq. (\ref{ComplexSEq}) and Eq. (\ref{ComplexSBC}) transform into
\begin{eqnarray}
\frac{\partial^2 u}{\partial t^2}
-v^2\frac{\partial^2 u}{\partial x^2}-\frac{1}{4\tau^2}u=0,
\label{ComplexuEq}\\
 \left.\frac{\partial u}{\partial x}\right|_\Gamma=0.
 \label{ComplexuBC}
\end{eqnarray}
The transformation given by Eq. (\ref{u}) is useful, in particular, since it allows us writing down the Green's function $S_G(x,t)$ of Eq. (\ref{ComplexSEq}) through the known Green's function of Eq. (\ref{ComplexuEq}) (we will take advantage of this fact below).

We note that Eq. (\ref{ComplexuEq}) coincides with the well-known Klein-Gordon equation (see, for example, Ref. \onlinecite{Bogoliubov}), but with an imaginary "mass" (see the sign of the last term). In relativistic quantum mechanics such an equation describes hypothetical particles called tachyons~\cite{Feinberg67a,Murphy72a,Recami86a,Dawe92a}. Because of the transformation (\ref{u}), the tachyon physics is somewhat 'hidden' in the dynamics of spin polarization and can be recovered by an inverse transformation.
Analogies between certain exotic particles and condensed matter systems have attracted a strong attention. For example, electrons in graphene~\cite{Novoselov05a} and graphite~\cite{Zhou06a} exhibit properties of Dirac fermions, signatures of magnetic monopoles were spotted in spin ice materials~\cite{Castelnovo08a}, and some exotic quasi-particle excitations in a variety of condensed-matter systems show a similarity with Majorana fermions~\cite{Wilczek09a}. Therefore, we believe that our observation that is exciting in itself will stimulate further theoretical and experimental work in this area.

\section{Spin relaxation in wires} \label{sec3}
In this section we present our studies of kinetics of spin relaxation in 1D infinite and finite length wires with Rashba spin-orbit coupling using the spin kinetic equation approach. Our analysis is essentially based on Eq. (\ref{ComplexSEq}) supplemented by initial and, where appropriate, by boundary conditions (\ref{ComplexSBC}). It is assumed that a thin narrow wire is made of a semiconductor heterostructure with spin-orbit coupling and the conduction electrons in the wire occupy the lowest size-quantization level corresponding to transverse confinement. In addition, the length of finite length wires is considered to be much longer than the phase coherence length. Therefore, the electron transport in the direction along the wire is described in terms of the classical transport regime.

\subsection{Relaxation of homogeneous polarization in infinite wires}
Let us start the analysis of solutions of Eq. (\ref{ComplexSEq}) with the most simple case, namely, with a problem of relaxation of homogeneous initial spin polarization in an infinite wire, $-\infty<x<\infty$. In this case, assuming that solutions of Eq. (\ref{ComplexSEq}) do not depend on $x$, we
can rewrite this equation and initial conditions as
\begin{eqnarray}
\frac{d^2 S}{d t^2}&+&\frac{1}{\tau}\frac{d S}{d t}+\Omega^2 S=0,
\label{SHomEq} \\
S(t=0)&\equiv & S(0)=S_x(0)+iS_z(0) \; , \; \left. \frac{d S}{d t}\right|_{t=0}=0.\;\;\; \label{eq3}
\end{eqnarray}
The solution of Eq. (\ref{SHomEq}) with initial conditions (\ref{eq3}) can be represented in the form
\begin{eqnarray}
S(t)=S(0)
e^{-\frac{t}{2\tau}}\left(
\cosh\kappa t+
\frac{\sinh\kappa t}{2\tau\kappa}\right),
\label{SHom}
\end{eqnarray}
where $\kappa=\sqrt{1/(4\tau^2)-\Omega^2}$. It should be emphasized that the parameter $\kappa$ can take both
real and imaginary values depending on system parameters. Moreover, in addition to describing spin polarization decay in infinite wires, Eq. (\ref{SHom}) also describes spin relaxation at long distances from boundaries of finite length (and semi-infinite) wires.

Generally, we can distinguish two spin relaxation regimes. It follows from Eq. (\ref{SHom}) that when $2\tau\Omega<1$ ($\textnormal{Im}(\kappa)=0$), the spin relaxation is described by two exponents with relaxation rates $1/(2\tau )\pm \kappa$.
In the limit of a small spin precession angle per mean free path, $\tau\Omega\ll 1$, we find
\begin{equation}
S(t)=S(0)
e^{-\tau\Omega^2 t}.
\label{SHomDif}
\end{equation}
The above expression coincides with the spin relaxation time predicted by D'yakonov-Perel' theory for 1D relaxation. It is clearly seen that in this situation the relaxation of spin polarization is characterized by the time constant
$(\tau\Omega^2)^{-1}$, which is much longer than $\tau$.

\begin{figure}[t]
 \begin{center}
\includegraphics[angle=0,width=8.0cm]{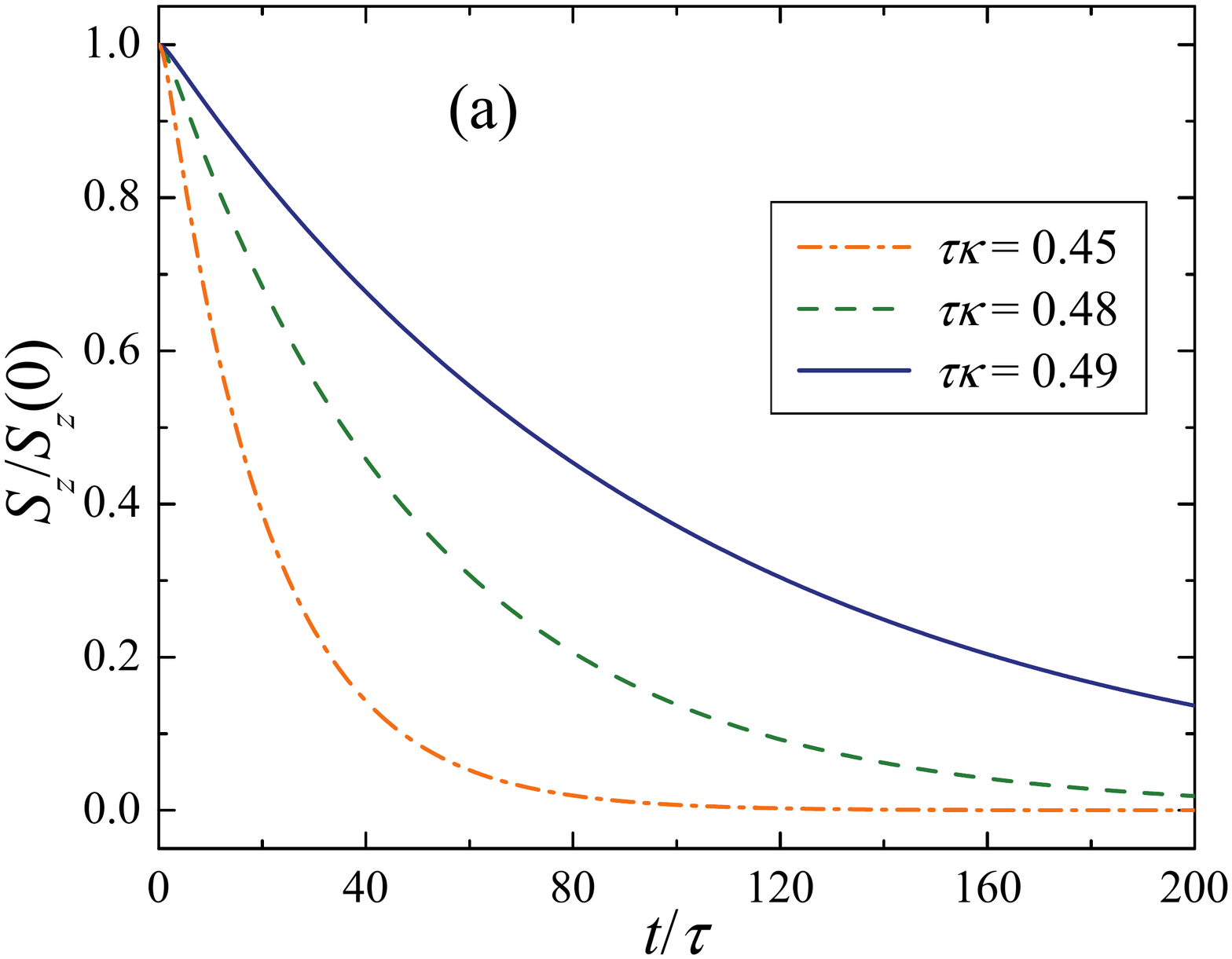}
\includegraphics[angle=0,width=8.0cm]{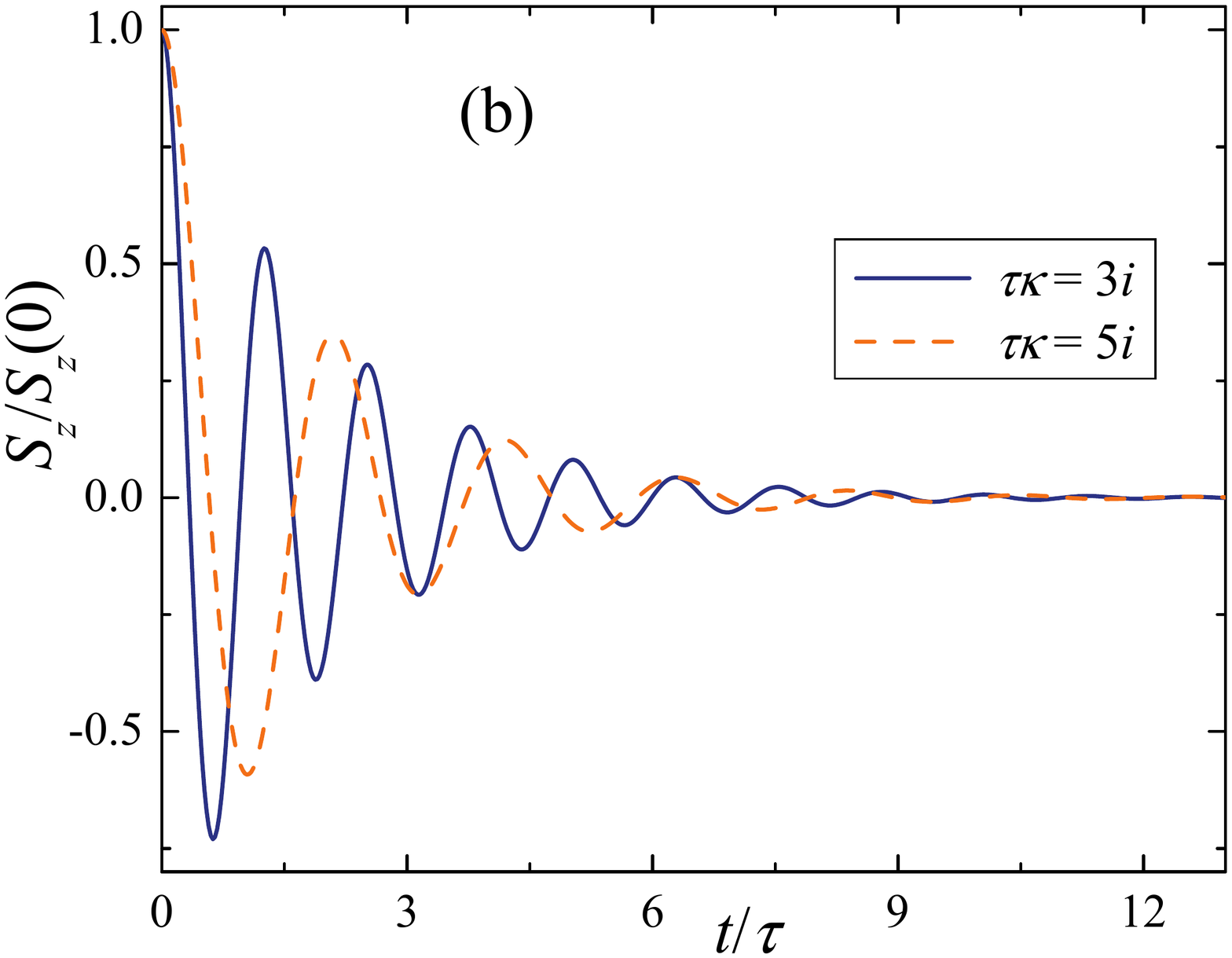}
\end{center}
\caption{(Color online) Dynamics of spin relaxation of homogeneous spin polarization in infinite wires. (a) and (b) correspond to two different regimes of spin relaxation of Eq. (\ref{SHom}) as discussed in the text. The values of $\kappa$ are indicated on the plots.} \label{fig1}
\end{figure}

The second regime of spin relaxation is realized when $2\tau\Omega>1$. In this case $\kappa$ is purely imaginary meaning
that the spin relaxation decay described by Eq. (\ref{SHom}) consists of an exponential decay with a rate of $1/(2\tau)$ modulated by oscillating
functions. In the limiting case $\tau\Omega\gg 1$ we obtain
\begin{eqnarray}
S(t)=S(0)
e^{-t/(2\tau)}\cos\Omega t.
\label{SHomBal}
\end{eqnarray}
Both regimes of spin relaxation are presented in Fig. \ref{fig1}.

\subsection{Relaxation of inhomogeneous polarization in infinite wires}

As it is mentioned below Eq. (\ref{ComplexuEq}), the Green's function of Eq. (\ref{ComplexSEq}), $S_G(x,t)$, can be obtained performing
a back transformation (from $u$ to $S$) of the known~\cite{VladimirovProblems} Green's function of Eq. (\ref{ComplexuEq})
(the forward transformation is given by Eq. (\ref{u})). Following this procedure we find
\begin{eqnarray}
S_G(x,t)=\frac{1}{2v}\Phi(vt-|x|)e^{i\eta x-\frac{t}{2\tau}}
I_0\left(\frac{\sqrt{v^2t^2-x^2}}{2v\tau}\right),
\label{SG}
\end{eqnarray}
where $\Phi(t)$ is the Heaviside step function, and $I_0(t)$ is the modified Bessel function
of zero order. We note that the Green's function (\ref{SG}) of Eq. (\ref{ComplexSEq}) describes the evolution of a point excitation
of spin polarization with the following initial conditions
\begin{eqnarray}
S_G(x,t)=0~
\textrm{for}~ t\leq 0, ~~
\left. \frac{\partial S_G}{\partial t}\right|_{t=0}=\delta(x).
\label{SGIC}
\end{eqnarray}
The Green's function (\ref{SG}) can be employed to determine the spin polarization at any moment of time for any given initial conditions using the relation
\begin{eqnarray}
S(x,t)=\left[\frac{\partial}{\partial t}+\frac{1}{\tau}\right]\int\limits_{-\infty}^{\infty}\textnormal{d}\xi S_{G}(x-\xi,t)S(\xi,0)\nonumber\\
+\int\limits_{-\infty}^{\infty}\textnormal{d}\xi S_{G}(x-\xi,t)\dot{S}(\xi,0).
\label{GenSol2}
\end{eqnarray}

In order to better understand the meaning of Eq. (\ref{SG}), let us consider the spin dynamics of the following initial excitation of spin polarization in $z$ direction:
\begin{equation}
S(x,t=0)=0 \; , \; \left. \frac{\partial S(x,t)}{\partial t}\right|_{t=0}=i\delta(x). \label{eq4}
\end{equation}
Accordingly to Eq. (\ref{GenSol2}), at any $t \geq 0$ the spin polarization in the system is given by
\begin{eqnarray}
S_x(x,t)&=&-\frac{\Phi(vt-|x|)}{2v}\sin \eta x e^{-\frac{t}{2\tau}}
I_0\left(\frac{\sqrt{v^2t^2-x^2}}{2v\tau}\right), \;\;\;\;\;
\label{SGx} ~\\
S_z(x,t)&=&\frac{\Phi(vt-|x|)}{2v}\cos \eta x e^{-\frac{t}{2\tau}}
I_0\left(\frac{\sqrt{v^2t^2-x^2}}{2v\tau}\right).\;\;\;\;\;
\label{SGz}
\end{eqnarray}
This solution describes a propagation of the initial excitation of spin polarization in both directions from the excitation point with well defined fronts.
Such a propagation also involves the spin precession and spin relaxation.

\begin{figure}[t]
 \begin{center}
\includegraphics[angle=0,width=8.0cm]{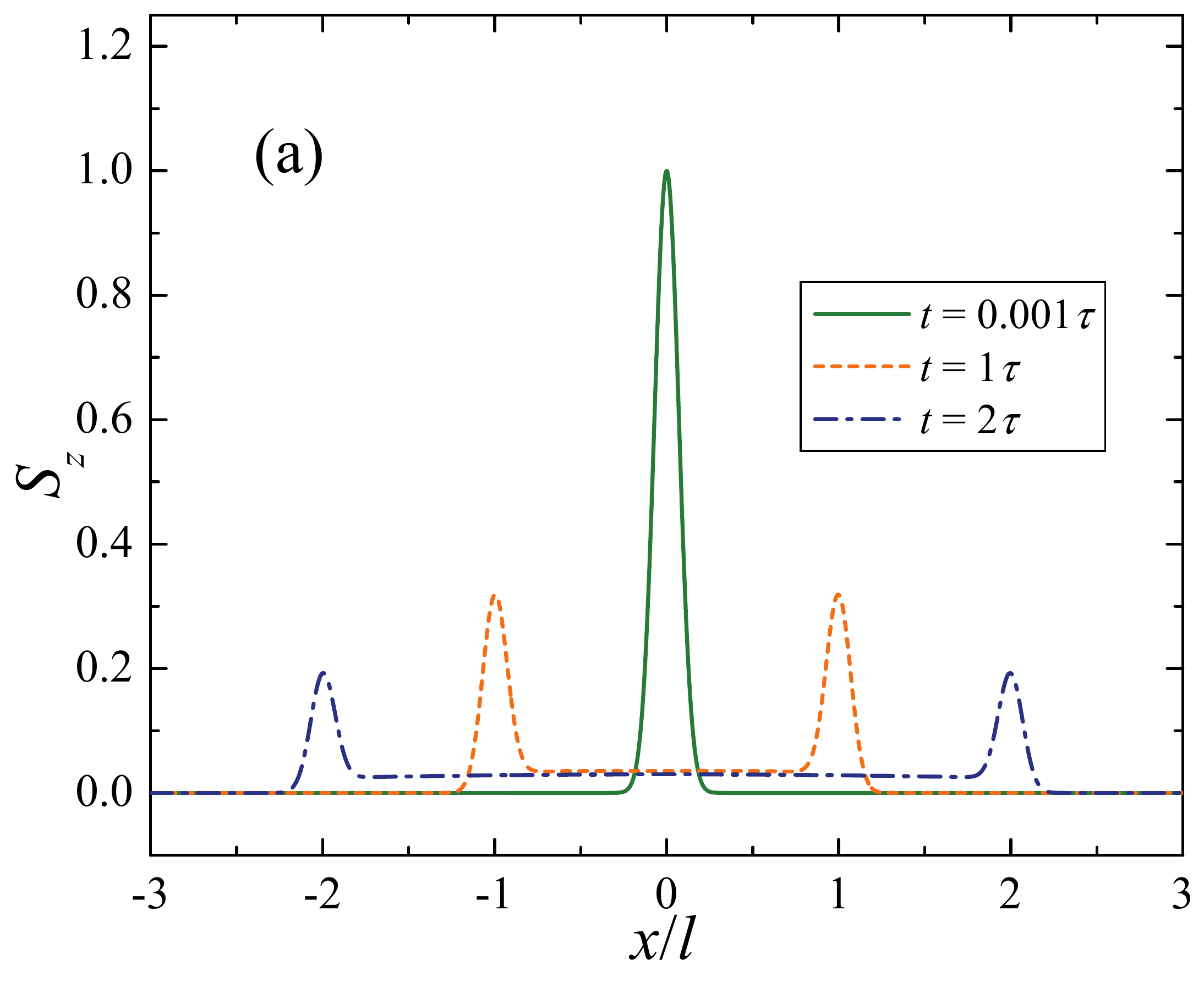}
\includegraphics[angle=0,width=8.0cm]{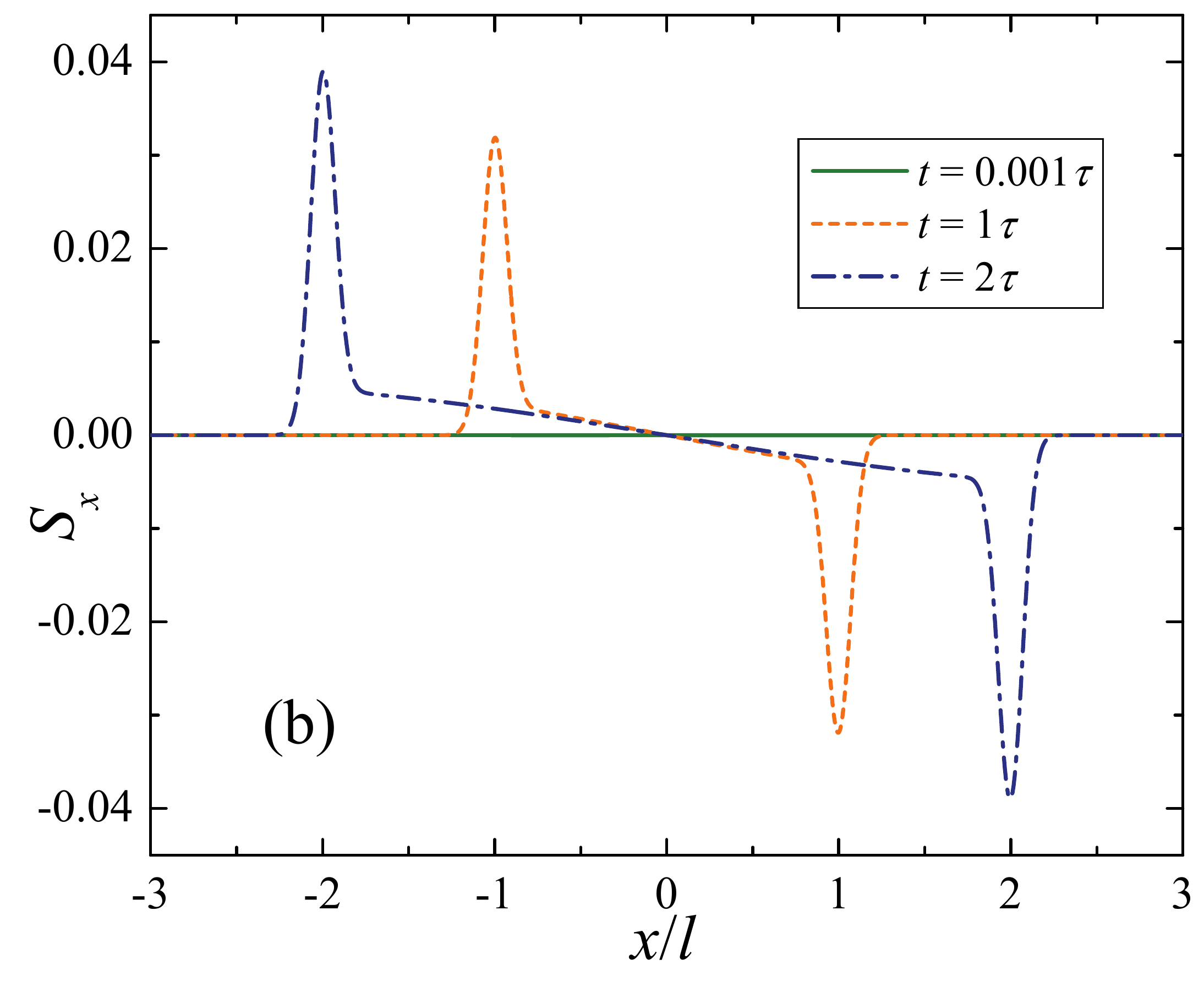}
\end{center}
\caption{(Color online) Dynamics of $S_z$ (a) and $S_x$ (b) components of spin polarization of an initial Gaussian spin polarization profile with spin polarization pointing in $z$ direction ($S(x,t=0)=i\textnormal{exp}(-100x^2/l^2)$). The spin polarization components are calculated using Eq. (\ref{GenSol2}). This plot was obtained using the parameter value $\eta l=0.1.$} \label{fig2}
\end{figure}

Fig. \ref{fig2} shows the dynamics of spin polarization in an infinite wire found using Eq. (\ref{GenSol2})
with different initial conditions. Specifically, we assume a Gaussian spin polarization profile and zero time derivative of spin polarization at $t=0$. This plot shows that the initial spin polarization profile splits into left- and right-moving packets whose amplitude in $z$ direction decreases in time. At the same time, the amplitude of $S_x$ in packets increases because of the spin precession. It should be emphasized that such moving packets of spin polarization are not captured by the drift-diffusion schemes of spin transport~\cite{Yu02a,Burkov04a,Saikin04a,Pershin04b,arXiv_1007_0853v1}.  In addition, we note that an area with a flat $S_z$ establishes between the moving packets. This region of flat distribution of $S_z$ can be considered as a precursor of diffusive dynamics.

Moreover, the expression for the Green's function (\ref{SG}) takes a simpler form at some specific points. For example, if we consider the position of the moving front, then
it is easy to obtain
\begin{eqnarray}
S_G(x=vt-0,t)=\frac{1}{2v}e^{i\eta x}e^{-\frac{t}{2\tau}}.
\label{SGFront}
\end{eqnarray}
This expression basically means that the amplitude of the moving front exponentially decreases with the time constant $2 \tau$. Another interesting point is $x=0$. We find that at $x=0$ the Green's function (\ref{SG}) is a real function of time:
\begin{eqnarray}
S_G(0,t)=\frac{1}{2v}e^{-\frac{t}{2\tau}}
I_0\left(\frac{t}{2\tau}\right).
\label{SG0}
\end{eqnarray}
Its asymptotic values at short and long times can be calculated. In particular, at short times $t\ll\tau$
\begin{eqnarray}
S_G(0,t)=\frac{1}{2v}
\left(
1-\frac{1}{2}\frac{t}{\tau}+\frac{3}{16}\frac{t^2}{\tau^2}+O\left(\frac{t^3}{\tau^3}\right)
\right),
\label{SG0Small}
\end{eqnarray}
 and at long times $t\gg\tau$
\begin{eqnarray}
S_G(0,t)=\frac{1}{2v}
\sqrt{\frac{\tau}{\pi t}}
\left(
1+O\left(\frac{\tau}{t}\right)
\right),
\label{SG0Large}
\end{eqnarray}
which corresponds to the well-known diffusive behavior.

\subsection{Relaxation of homogeneous polarization in finite length wires} \label{hfl}

Let us consider the problem of spin relaxation in wires of finite length $L$, $0<x<L$, when at the initial moment of time the electron spins are homogeneously polarized along $z$-axis,
\begin{eqnarray}
S(x,0)=iS_0\; , \; \left. \frac{\partial S}{\partial t}\right|_{t=0}=0.
\label{HomSIC}
\end{eqnarray}
We find the solution of Eq. (\ref{ComplexSEq}) with the boundary conditions (\ref{ComplexSBC}) and initial conditions (\ref{HomSIC}) using the standard method of separation of variables. A straightforward application of this method leads to the following expression for the complex spin polarization
\begin{eqnarray}
\frac{S(x,t)}{S_0}=i\frac{\sin(\eta L/2)}{\eta L/2}e^{i\eta ( x-L/2)}+2\eta L e^{i\eta x-t/(2\tau)} \nonumber \\
 \times\sum_{n=1}^{+\infty}\frac{1-(-1)^n
e^{-i\eta L}}{(\eta L)^2-(\pi n)^2}
\left(
\cosh\kappa_n t+
\frac{\sinh\kappa_n t}{2\tau\kappa_n}\right)
\cos\frac{\pi n x}{L}, \;\;
\label{SHomSol}
\end{eqnarray}
where $\kappa_n=\sqrt{1/(4\tau^2)-\pi^2 n^2v^2/L^2}$.

\begin{figure}[t]
 \begin{center}
\includegraphics[angle=0,width=8.0cm]{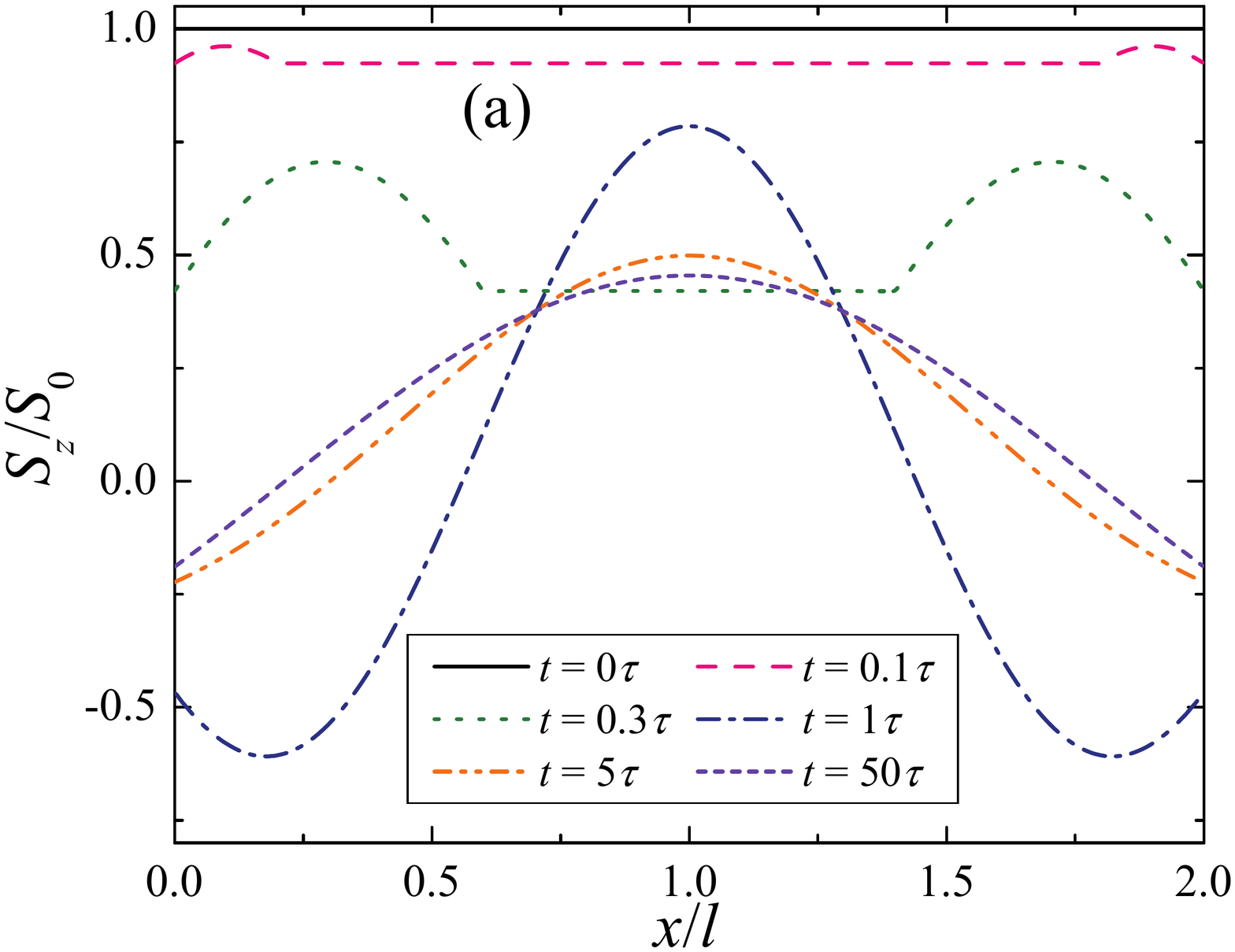}
\includegraphics[angle=0,width=8.0cm]{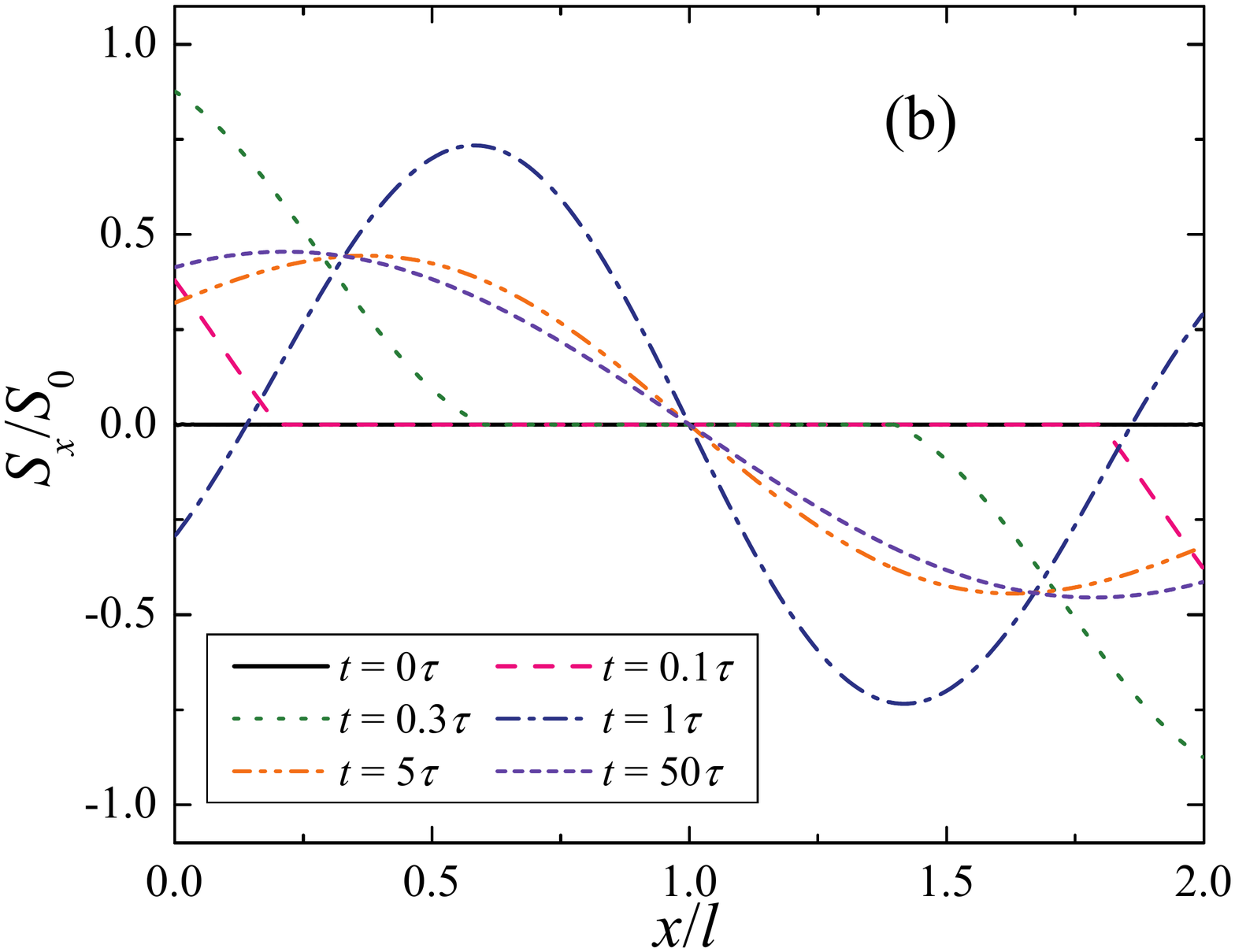}
\end{center}
 \caption{(Color online) Dynamics of formation of persistent spin helix in the ballistic regime of spin transport. This plot was obtained using the parameter values $\eta l=2$ and $L=2l.$ \label{fig3}}
\end{figure}

Previously, we investigated the spin relaxation in finite length wires in the diffusive regime and found that an initially homogeneous spin polarization in $z$ direction transforms into a persistent spin helix at long times~\cite{Slipko11a}. The same qualitative result follows from Eq. (\ref{SHomSol}): note that only the first term in the right-hand side of Eq. (\ref{SHomSol}) survives at long times. However, Eq. (\ref{SHomSol}) by itself is more complex compared to those reported in Ref. \onlinecite{Slipko11a} for the case of diffusive spin transport because it incorporates both diffusive and ballistic spin transport. It can be shown that Eq. (\ref{SHomSol}) gives the same result~\cite{Slipko11a} in the diffusive limit. Indeed,
when $\ell=v\tau\ll L$ and $t\gg\tau$ it is easy to demonstrate that
\begin{eqnarray}
\lim_{\tau\rightarrow 0, \ell\rightarrow 0} e^{-t/(2\tau)}\left(
\cosh\kappa_n t+
\frac{\sinh\kappa_n t}{2\tau\kappa_n}\right)=
e^{-\frac{\pi^2 n^2 Dt}{L^2}}, \;\;
\label{limit}
\end{eqnarray}
where $ D=\ell^2/\tau$. Substituting Eq. (\ref{limit}) into Eq. (\ref{SHomSol}) we obtain the same expression for $S(x,t)$ as
in Ref. \onlinecite{Slipko11a}. The dynamics of formation of persistent spin helix in the ballistic regime of spin transport is depicted in Fig. \ref{fig3}. This figure is obtained plotting Eq. (\ref{SHomSol}) at different moments of time. One notices that the persistent spin helix is almost entirely formed within a short $\sim 5 \tau$ time interval.

Let us consider the  ballistic limit in more details. In this case assuming $\ell\gg L$, $2\kappa_n\tau/i \approx  2\pi n\ell/L\gg 1$
the second term in the big round brackets in Eq. (\ref{SHomSol}) can be omitted (since it involves a small multiplier $1/(2\kappa_n\tau)$). Then, we obtain
%\begin{eqnarray}
%\frac{S(x,t)}{S_0}=i\frac{\sin(\eta L/2)}{\eta L/2}e^{i\eta ( x-L/2)}+\eta L e^{i\eta x-t/(2\tau)} \nonumber \\
% \times\sum_{n=1}^{+\infty}\frac{1-(-1)^n
%e^{-i\eta L}}{(\eta L)^2-(\pi n)^2}
%\left(\cos\frac{\pi n (x+vt)}{L}+\cos\frac{\pi n (x-vt)}{L}\right).~ \;\;
%\label{SHomSolB}
%\end{eqnarray}
\begin{widetext}
\begin{equation}
\frac{S(x,t)}{S_0}=i\frac{\sin(\eta L/2)}{\eta L/2}e^{i\eta ( x-L/2)}+\eta L e^{i\eta x-t/(2\tau)} 
\sum_{n=1}^{+\infty}\frac{1-(-1)^n
e^{-i\eta L}}{(\eta L)^2-(\pi n)^2}
\left(\cos\frac{\pi n (x+vt)}{L}+\cos\frac{\pi n (x-vt)}{L}\right).~ \;\;
\label{SHomSolB}
\end{equation}
\end{widetext}
We can sum the series in Eq. (\ref{SHomSolB}) taking into account the fact that at the initial moment of time, $t=0$, the right-hand side of Eq. (\ref{SHomSolB}) is equal to $i$. Moreover, since these Fourier series are even and $2L$-periodic, we can finally obtain  a very simple expression for the spin polarization in  the ballistic limit:
\begin{eqnarray}
\frac{S(x,t)}{S_0}=i\frac{\sin(\eta L/2)}{\eta L/2}e^{i\eta ( x-L/2)}\left(1-e^{-t/(2\tau)}\right) \nonumber \\
+\frac{i}{2} e^{-t/(2\tau)}\left(  e^{i\eta(x-|\widetilde{x-vt}|)}+ e^{i\eta(x-|\widetilde{x+vt}|)} \right), \;\;
\label{SHomSolBF}
\end{eqnarray}
where $\widetilde{z}=z+2Ln$, and $n$ is an integer number selected in such a way that for any real $z$, $-L<\widetilde{z}\leq L$.

\subsection{Relaxation of inhomogeneous polarization in finite length wires} \label{sec34}

Next, we investigate the dynamics of relaxation of an arbitrary spin polarization profile in finite length wires. For this purpose,
we obtain an expression for the Green's function $G(x,t;\xi)$ that allows finding the spin polarization profile at any moment of time for given initial conditions. The Green's function $G(x,t;\xi)$ is calculated by mapping Eq. (\ref{u}) into Eq. (\ref{ComplexuEq}) taking into account the boundary conditions (\ref{ComplexuBC}) by an even mapping of the initial condition on an infinite wire.
It is straightforward to show that $G(x,t;\xi)$ can be written as
\begin{eqnarray}
G(x,t;\xi)=\frac{e^{i\eta(x-\xi)-t/(2\tau)}}{2v}\sum_{n=-\infty}^{+\infty}
\left[
\Phi(vt-|x-\xi-2Ln|)\right.\nonumber\\
\times I_0\left( \frac{\sqrt{v^2t^2-(x-\xi-2Ln)^2}}{2v\tau} \right)
+\nonumber\\
\left.\Phi(vt-|x+\xi-2Ln|)
I_0\left( \frac{\sqrt{v^2t^2-(x+\xi-2Ln)^2}}{2v\tau} \right)
\right].~~
\label{G}
\end{eqnarray}
The spin polarization distribution along the wire at any moment of time can be calculated using the following equation:
\begin{eqnarray}
S(x,t)=\left[\frac{\partial}{\partial t}+\frac{1}{\tau} \right]\int_{0}^{L}d\xi G(x,t;\xi)S(\xi,0)\nonumber\\
+\int_{0}^{L}d\xi G(x,t;\xi)\dot{S}(\xi,0).
\label{GenSolS}
\end{eqnarray}

\begin{figure}[t]
 \begin{center}
\includegraphics[angle=0,width=8.0cm]{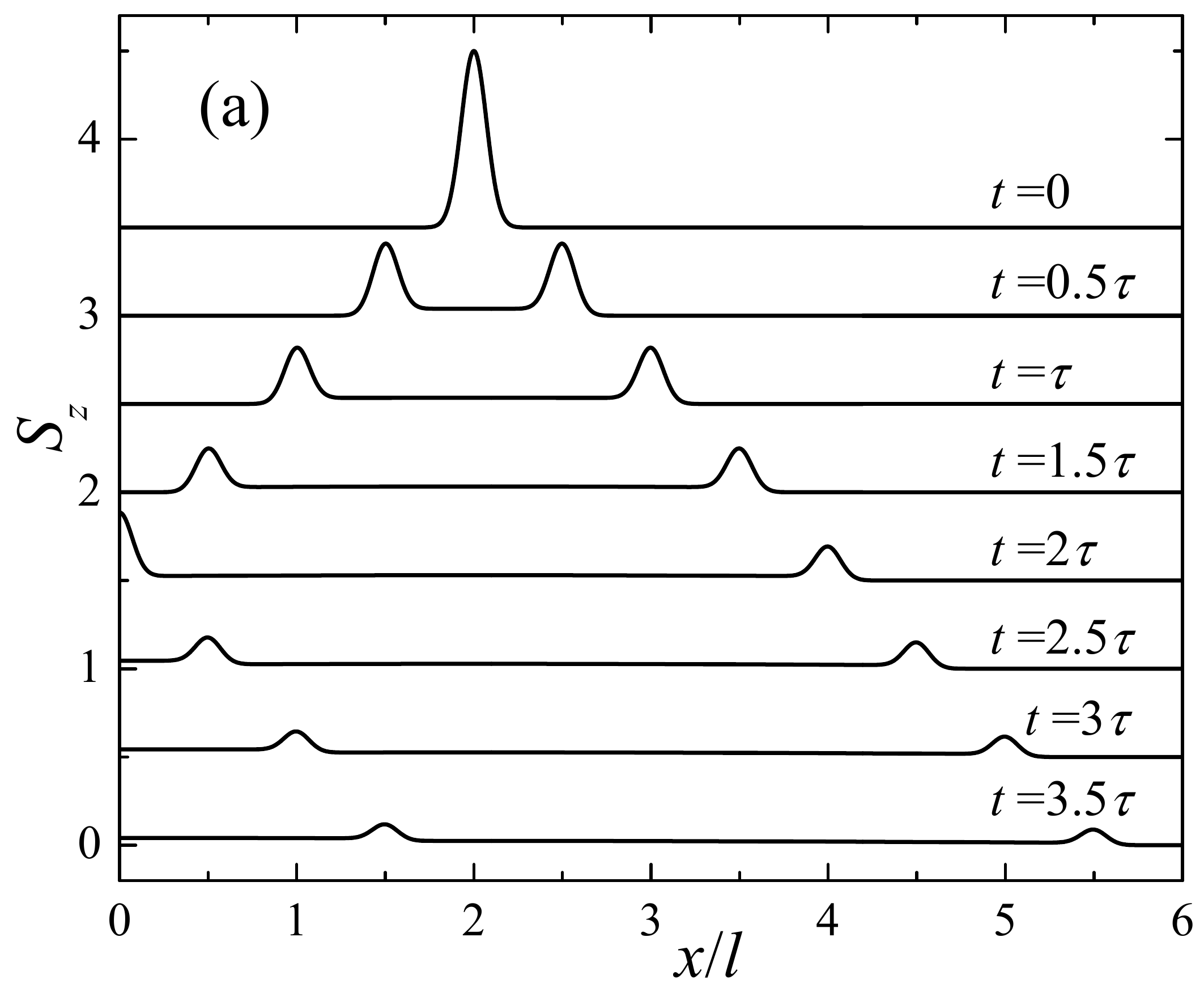}
\includegraphics[angle=0,width=8.0cm]{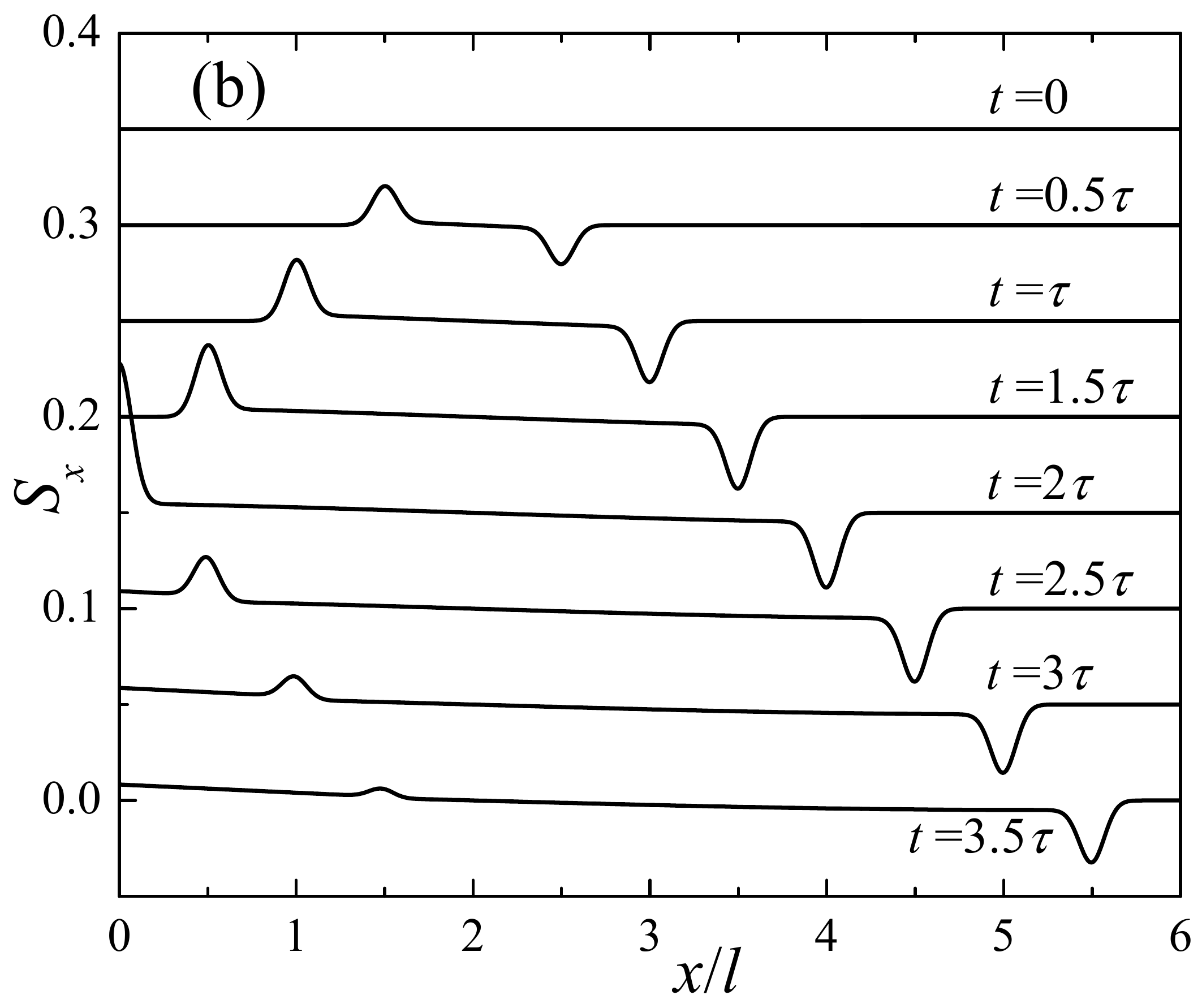}
\end{center}
 \caption{(Color online) The boundary spin echo in a finite length wire. The evolution of an initially Gaussian spin polarization profile pointing in $z$ direction ($S(x,t=0)=i\textnormal{exp}(-100x^2/l^2)$) was found using Eqs. (\ref{G}) and (\ref{GenSolS}). It is clearly seen that the spin polarization packet initially moving in $-z$ direction moves back after the reflection from the boundary. This plot was obtained using the parameter values $\eta l=0.1$, $L=6l$. The curves are displaced for clarity. \label{fig4}}
\end{figure}

As an example of spin dynamics calculation in finite length wires, let us consider the evolution of a Gaussian initial spin polarization profile. The time derivative of spin polarization at $t=0$ is set equal to zero (this is the second initial condition). Fig. \ref{fig4} shows results of our calculations based on Eqs. (\ref{G}) and (\ref{GenSolS}). Specifically, it is clearly seen that the initial spin polarization profile splits into two packets moving in opposite direction similarly to the result shown in Fig. \ref{fig2} for the case of infinite wire. The amplitude of $S_z$ in these packets decreases in time. At $t=2\tau$ the left-moving packet reaches the wire boundary and is reflected back changing its direction of motion. We call such a reflected packet of spin polarization as {\it the boundary spin echo} to distinguish it from the spin echo in nuclear magnetic resonance (NMR)~\cite{Abragam83}. The boundary spin echo is indeed much closer reassembles the echo of sound waves than those in NMR.  We also note that at long times any initial spin polarization profile (including one shown in Fig. \ref{fig4}) shapes into the persistent spin helix characterized by a certain amplitude and phase.

\section{Spin relaxation in 2D channels} \label{sec4}

Let us consider a two-dimensional electron gas in the presence of linear in the wave vector Rashba and Dresselhaus spin-orbit couplings.
The electron Hamiltonian is written as
\begin{equation}
 H=\frac{\mathbf{ p}^2}{2m}+\alpha\ (p_y\sigma_x-p_x\sigma_y)+\beta (p_x\sigma_x-p_y\sigma_y),
\label{ham2D}
\end{equation}
where $\alpha$ and $\beta$ are strengths of Rashba and Dresselhauss spin-orbit couplings correspondingly. In the specific case that we consider below, namely
when $\alpha=\beta$, the Hamiltonian (\ref{ham2D}) takes the form
\begin{equation}
 H=\frac{\mathbf{ p}^2}{2m}+ \alpha  (\sigma_x-\sigma_y)(p_x+p_y).
\label{ham2Da}
\end{equation}
Next, consider a rotation about $z$ axis by an angle $\varphi=\pi/4$. Then, $x$ is transformed into $x'=(x+y)/\sqrt{2}$ and $y'=(y-x)/\sqrt{2}$.
The transformation of the spin part of the Hamiltonian (\ref{ham2Da}) is performed using the standard unitary transformation
\begin{equation}
U=\left(
\begin{array}{cc}
e^{i\varphi/2} & 0  \\
0 & e^{-i\varphi/2}  \end{array}
\right).
\end{equation}
In the new basis the Hamiltonian (\ref{ham2Da}) takes the form
\begin{equation}
 {H}'=U {H} U^\dagger=
\frac{(\mathbf{p})^2}{2m}-2 \alpha p_{x'} \sigma'_{y'},
\label{ham2Db}
\end{equation}
where $\sigma'_{y'}=\sigma_{y}$ corresponds to the projection of spin operator in the rotated basis on $y'$.
Under the action of the Hamiltonian $H'$, the spin of an electron precesses with an angular velocity
$\mathbf{\Omega'_p}=-\Omega'\mathbf{e}_{y'} \cos\theta $, where $\theta$ is the angle between the direction of electron momentum and $x'$ axis
and $\Omega'=4\alpha p/\hbar$.

Now, let us consider electron spin dynamics in a channel of width $L$ in $x'$ direction and of infinite length in $y'$ direction. Our goal is to show that the evolution of spin polarization (homogeneous in $y'$ direction) in such a channel is similar to the evolution of  spin polarization in finite length wires considered in Sec. \ref{sec3}. Assuming that the spin polarization in the channel depends only on $x'$ and $S_{y'}=0$, we can combine Eqs. (\ref{KinEq}) and (\ref{CollisionIntegral}) as
\begin{eqnarray}
\frac{\partial \mathbf{S}(\theta)}{\partial t}+v \cos\theta \frac{\partial \mathbf{S}(\theta)}{\partial x'}=-\Omega'\cos\theta \mathbf{e}_{y'}\times\mathbf{S}(\theta)\nonumber\\
-\frac{1}{\tau}
\left[
\mathbf{S}(\theta)-\int_{0}^{2\pi}\frac{d\theta}{2\pi}\mathbf{S}(\theta)
\right],
\label{S2DIntEq}
\end{eqnarray}
where the following shorthand notation is used: $\mathbf{S}(\theta)=\mathbf{S}(\theta,x',t)$. Introducing a complex spin polarization $S(\theta)=S_{x'}(\theta)+iS_z(\theta)$,
we arrive to the following integral equation
\begin{eqnarray}
\frac{\partial S(\theta)}{\partial t}+v \cos\theta \frac{\partial S(\theta)}{\partial x}=i\Omega\cos\theta S(\theta)\nonumber\\
-\frac{1}{\tau}
\left[
S(\theta)-\int_{0}^{2\pi}\frac{d\theta}{2\pi}S(\theta)
\right]
\label{CS2DIntEq}
\end{eqnarray}
complimented by the boundary condition
\begin{eqnarray}
\left.(S(\theta)-S(\pi-\theta))\right| _\Gamma=0.
\label{CS2DIntBC}
\end{eqnarray}
We seek the solution of Eq. (\ref{CS2DIntEq}) in the form of Fourier series
\begin{eqnarray}
S(\theta)=\sigma_0+\sum_{n=1}^{+\infty}(\sigma_n\cos n\theta+\bar{\sigma}_n\sin n\theta),
\label{CSseries}
\end{eqnarray}
where $\sigma_n$ and $\bar{\sigma}_n$ are functions of $x'$ and $t$. In particular, it is clear that $\sigma_0=\int_{0}^{2\pi}d\theta S(\theta)/(2\pi)$ is
proportional to the spin polarization.

Substituting Eq. (\ref{CSseries}) into Eq. (\ref{CS2DIntEq}) and using the completeness of the set of function
$\{ 1, \cos n\theta, \sin n \theta \}$ we find infinite series of interconnected equations. The two first equations in the series are:
\begin{eqnarray}
\frac{\partial \sigma_0}{\partial t}+\frac{1}{2}\left(v\frac{\partial}{\partial x'}-i\Omega'\right)\sigma_1=0,
\label{Sigma0Eq}
\\
\frac{\partial \sigma_1}{\partial t}+\left(v\frac{\partial}{\partial x'}-i\Omega'\right)\left(\sigma_0+\frac{\sigma_2}{2}\right)+\frac{\sigma_1}{\tau}=0.
\label{Sigma1Eq}
\end{eqnarray}
For a weakly anisotropic electron momentum distribution $|\sigma_2|\ll|\sigma_0|$, therefore, we can truncate the series by setting $\sigma_2=0$ in Eq.
(\ref{Sigma1Eq}). Thus, Eq. (\ref{Sigma1Eq}) can be rewritten as
\begin{eqnarray}
\frac{\partial \sigma_1}{\partial t}+\left(v\frac{\partial}{\partial x'}-i\Omega'\right)\sigma_0+\frac{\sigma_1}{\tau}=0
\label{Sigma1FEq}
\end{eqnarray}
Excluding $\sigma_1$ from Eqs. (\ref{Sigma0Eq}), (\ref{Sigma1FEq}), we obtain the following equation for the total spin polarization $\sigma_0$:
\begin{eqnarray}
\frac{\partial^2 \sigma_0}{\partial t^2}+\frac{1}{\tau}\frac{\partial \sigma_0}{\partial t}
-\left(\frac{v}{\sqrt{2}}\frac{\partial}{\partial x'}-i\frac{\Omega'}{\sqrt{2}}\right)^2 \sigma_0=0
\label{Sigma0FEq}
\end{eqnarray}
which is of the similar form with Eq.(\ref{ComplexSEq}). This is the main result of Sec. \ref{sec4} proving similarity of spin relaxation in finite length wires and 2D channels.

The boundary condition for spin dynamics in the channel can be obtained in the following way. Substituting the approximate solution of Eq. (\ref{CS2DIntEq}),
\begin{eqnarray}
S(\theta)=\sigma_0+\sigma_1\cos\theta +\bar{\sigma}_1\sin\theta,
\label{SThetaSol}
\end{eqnarray}
into the boundary condition Eq. (\ref{CS2DIntBC}), we immediately notice that the function $\sigma_1$ on the channel boundary must turn to zero, namely
\begin{eqnarray}
\left. \sigma_1\right|_{\Gamma}=0.
\label{Sigma1BC}
\end{eqnarray}
Combining Eq. (\ref{Sigma1BC}) with (\ref{Sigma1FEq}), we find the corresponding  boundary condition
for the complex spin polarization $\sigma_0$,
\begin{eqnarray}
 \left.\left(\frac{\partial \sigma_0}{\partial x'}-i\frac{\Omega '}{v} \sigma_0\right)\right|_\Gamma=0,
 \label{Sigma0BC}
\end{eqnarray}
which is similar to the boundary condition in finite length wires given by Eq. (\ref{ComplexSBC}).

\section{Conclusions} \label{sec5}

In this paper, we studied spin relaxation in 1D and 2D systems with spin-orbit interaction. For this purpose,
we used the spin kinetic equation that takes into account both ballistic and diffusive spin transport regimes. This formalism was applied
to several interesting problems of spin relaxation that were solved analytically. In particular, we analyzed dynamics of homogeneous and inhomogeneous spin
polarizations in infinite and finite length 1D wires. Moreover, it was explicitly demonstrated that the problem of spin relaxation in appropriately oriented 2D channels with Rashba and Dresselhaus interactions of equal strength can be reduced to the problem of spin relaxation in 1D wires. This result establishes a solid foundation for a recently suggested method of creation of persistent spin helical configurations in semiconductors~\cite{Slipko11a}.

One of the main results of the paper is a prediction of the spin echo in the ballistic regime of spin transport. We found that in finite length wires an initially localized spin polarization profile reflects from a sample boundary returning to its initial position. To distinguish such a spin echo from that in NMR and emphasize its closer analogy with the spin echo of sound waves we named such an effect as "the boundary spin echo". Another interesting result is a discovery of a transformation that maps the spin kinetic equation into the Klein-Gordon equation with an imaginary mass. In the relativistic quantum mechanics, this equation is used to describe tachyons, namely, hypothetical subatomic particles that moves faster than light. Therefore, we believe that certain predictions  of relativistic quantum mechanics can be laboratory tested. Of course, much work is to be done in this direction.

\vspace{1cm}

\section*{Acknowledgment}
We thank Professor M. Di Ventra for an interesting discussion.

\bibliography{spin}

\end{document}